\documentstyle[preprint,prl,aps]{revtex}
\input{epsf}

\begin{document}
\draft
\title{
TRANSMISSION THROUGH AN INTERACTING QUANTUM DOT\\
IN THE COULOMB BLOCKADE REGIME}

\author
{R. Berkovits$^1$, U. Sivan$^2$}

\address{
$^1$The Jack and Pearl Resnick Institute of Advanced Technology,\\
Department of Physics, Bar-Ilan University,
Ramat-Gan 52900, Israel}

\address{
$^2$Solid State Institute and Physics Department, Technion,\\
Israel Institute of Technology, Haifa, Israel}

\date{\today}
\maketitle

\begin{abstract}
The influence of electron-electron (e-e) interactions on the 
transmission through a quantum dot is investigated numerically 
for the Coulomb blockade regime. For vanishing magnetic fields,
the conductance peak height statistics is found to be independent of
the interactions strength. It is identical to the statistics predicted
by constant interaction single electron random matrix theory 
and agrees well with recent experiments.
However, in contrast to these 
random matrix theories, our calculations reproduces
the reduced sensitivity to magnetic flux observed in many experiments.
The relevant physics is traced to the short range Coulomb correlations
providing thus a unified explanation for the transmission statistics 
as well as for the large conductance peak spacing fluctuations observed in
other experiments.
\end{abstract}

Since the discovery of the Coulomb blockade phenomenon, most tunneling
experiments through a quantum dot were interpreted within the constant
interaction model (CI)\cite{Kastner,Meirav}. In that approximation, the
ground state energy of a quantum dot populated by $N$ electrons is expressed
as $E_N=\frac{e^2 N^2}{2C}+\sum_{i=1}^N\eta _i$ where $C$ is the dots
constant (or slowly varying) capacitance, and $\eta _i$ are the single
particle energies. Evidently, in that model only the long range Coulomb
interaction is taken into account while the short range correlations are
neglected. The fine ground state properties are hence determined by the
single particle states which for a disordered or chaotic dot display a
random matrix theory (RMT) statistics. Although the CI model is very
appealing in its simplicity, it was recently proved wrong in predicting the
the distribution of the conductance peak spacings\cite
{sba,shw,cm}, as well as the
large peak spacing fluctuations found in some of the experiment\cite
{sba,shw}. While the CI model predicts a RMT type of ground state
statistics with a characteristic energy scale, $\Delta $ (average single
particle level spacing), the experiments find a different type of statistics
and (at least for the experiments in Ref. \cite{sba,shw})
considerably larger
fluctuations which seem to be independent of $\Delta $. Moreover, the
ground state energy turns out to be relatively insensitive to magnetic flux
\cite{shw} and application of one quantum flux unit, $\phi _0=hc/e$,
through the dot hardly affect it\cite{marc,chang}. 
This insensitivity is again in
contrast with the CI model since the single particle states, and hence the
ground state energy in that model,
are expected to fluctuate on a flux scale smaller than
one quantum flux unit\cite{Simon,Sivan94}.

A point of view similar to the CI one was also taken for the calculation of
the Coulomb blockade conductance peak height statistics\cite
{jsa,pei,aa,blm}. Since RMT was assumed for the single particle
states, a modified Porter-Thomas distribution was obtained for 
the dimensionless
transmission $\alpha \equiv 2\Gamma _L\Gamma _R/\Gamma \langle \Gamma \rangle$ 
\begin{eqnarray}
P_{B=0}(\alpha )=\sqrt{\frac 2{\pi \alpha }}e^{-2\alpha }; 
\ \ \ \ \ \ \ \ \ \ \nonumber\\
P_{B\neq0}(\alpha )=4\alpha [K_0(2\alpha )+K_1(2\alpha )]e^{-2\alpha }.
\label{pa}
\end{eqnarray}
Here,
$\Gamma _L(\Gamma _R)$ are the tunneling rates from the left (right) lead,
$\Gamma =\Gamma _L+\Gamma _R$, $\langle \ldots \rangle$ denotes an average
over different peaks or disorder realizations and
$K_0,K_1$ are the modified Bessel functions. 

Subsequent experiments\cite{marc,chang} reported partial agreement with
these calculations and one was therefore facing the following dilemma: while
the conductance peak spacing fluctuations can be explained by short range
Coulomb correlations\cite{sba} 
(see also recent papers by Koulakov et al.\cite
{Koulakov} and Blanter et al.\cite{Blanter}), 
the conductance peak height statistics roughly agrees with a
model that totally neglects these correlations. It is hard to reconcile such
two different pieces of physics for two facets of the same phenomenon and in
the present manuscript we show indeed that the observed peak height statistics
may result from fluctuations in the short range Coulomb correlations
rather than the single electron RMT physics (i.e., wave functions with no 
correlations)
utilized in refs.\cite
{jsa,pei,aa,blm}. Strong support in favor of the Coulomb correlations type of
physics comes from the insensitivity to magnetic flux observed in both
experiments \cite{marc,chang}.
The auto correlation
function between the height of a given peak at two different values of the
magnetic flux, $\phi $,
\begin{equation}
C(\phi ,\Delta \phi )\equiv \frac{\left\langle \delta \alpha (\phi )\delta
\alpha (\phi +\Delta \phi )\right\rangle }{\sqrt{\left\langle \delta
^2\alpha (\phi )\right\rangle \left\langle \delta ^2\alpha (\phi +\Delta
\phi )\right\rangle }} \ ; \ \ \delta \alpha
=\alpha -\left\langle \alpha \right\rangle,
\label{acd}
\end{equation}
is found experimentally 
to decay on flux scales larger than predicted by CI single electron
RMT. This reduced sensitivity to flux is clearly
manifested in the numerical calculation presented below that take the 
interactions into account. Curiously, the inclusion of interactions
does not change the peak hight distribution predicted by single electron
RMT and observed by the experiments.
We are therefore able to propose a unified
explanation for both facets of the Coulomb blockade phenomenon, namely, the
conductance peak spacing fluctuations and the peak height
distribution. They both may originate from fluctuations in the Coulomb
interaction rather than single particle physics\cite{sba}. 

The height $g_{max}$ of a conductance peak is given by
\begin{eqnarray}
g_{max}={{e^2}\over{h}}\left({{\pi}\over{2k_B T}}\right) 
\langle{\Gamma}\rangle\alpha.
\label{gmax}
\end{eqnarray}
The tunneling rates may
be formulated in a tight-binding many particle language as
\begin{eqnarray}
\Gamma_{L (R)} = |t_{L (R)}|^2 |\sum_{k,j\in[L (R)]}
\langle \Psi_{N+1} | a_{k,j}^{\dag} |\Psi_{N} \rangle|^2,
\label{gamma}
\end{eqnarray}
where $t_{L (R)}$ is the barrier transmission which is assumed to
depend only weakly on energy, $\Psi_{N}$ is the $N$ particle
ground state wave function in the dot,
$a_{k,j}^{\dag}$ is the fermionic creation operator at site ($k,j$), and 
summation is performed on sites $k,j\in[L (R)]$ i.e, sites
adjacent to the 
left (right) lead. This is a straight forward adaptation of the
definition of $\Gamma_{L (R)}$ given in Ref. \cite{jsa}. For the
non-interacting case, assuming statistically identical independent 
single channel leads (for example, point contacts\cite{pei}), RMT
predicts the distribution depicted in Eq. (\ref{pa})\cite{jsa}.

We calculate the tunneling rates $\Gamma_{L (R)}$
for a system of interacting electrons modeled by 
a tight-binding Hamiltonian.
We choose a 2D cylindrical geometry of circumference 
$L_x$ and height $L_y$. This particular geometry is
very convenient for the study of the influence of a
magnetic flux $\phi$ threading the cylinder in the $\hat y$ direction.
A radial magnetic field could also be applied,
but for a field equivalent to one quantum flux unit through the
system, Landau bands appear. Since this is not the situation in the experiment
we prefer to apply a threading flux only.
The Hamiltonian is given by:
\begin{eqnarray}
H= \sum_{k,j} \epsilon_{k,j} a_{k,j}^{\dag} a_{k,j} - V \sum_{k,j}
(\exp(i 2 \pi (\phi/\phi_0) s/L_x)
a_{k,j+1}^{\dag} a_{k,j} + a_{k+1,j}^{\dag} a_{k,j} + h.c)
+ H_{int},
\label{hamil}
\end{eqnarray}
where  
$\epsilon_{k,j}$ is the energy of a site ($k,j$), chosen 
randomly between $-W/2$ and $W/2$ with uniform probability, $V$
is a constant hopping matrix element, and
$s$ is the lattice unit. The interaction
Hamiltonian is given by:
\begin{equation}
H_{int} = U  
\sum_{k,j>l,p} {{a_{k,j}^{\dag} a_{k,j}
a_{l,p}^{\dag} a_{l,p}} \over 
{|\vec r_{k,j} - \vec r_{l,p}|/s}}
\label{hamil2}
\end{equation}
where $U=e^2/s$. The distance $|\vec r_{k,j} - \vec r_{l,p}|/s=
(\min\{(k-l)^2,(L_x/s - (k-l))^2\}
+(j-p)^2)^{1/2}$.
The interaction term represents Coulomb
interaction between electrons confined to a 2D
cylinder embedded in a 3D space.

We consider a $4 \times 6$ cylinder with $M=24$ sites and $N=3$ or $N=4$
electrons. The size of the many body Hilbert space is 
$m = (_N^M)$. The $m \times m$ Hamiltonian
matrix is numerically
diagonalized and the ground-state eigenvectors $\Psi_N$ are obtained. 
The strength of e-e interactions, $U$, is varied between $0-22V$. 
The disorder strength is set to $W=3V$
in order to assure RMT behavior
for the non-interacting case. 
For each value of $U$, the results are averaged over
$500$ different realizations of disorder.
The left lead is attached to the ($1,1$) site
and the right one to the ($4,6$) site (point
contacts). Assuming a fixed barrier transmission $|t_{L (R)}|^2\equiv1$,
$\Gamma_{L (R)}$ is calculated using Eq. (\ref{gamma}).

\begin{figure}
\vspace{-1.5cm}
\centerline{\epsfxsize = 3in \epsffile{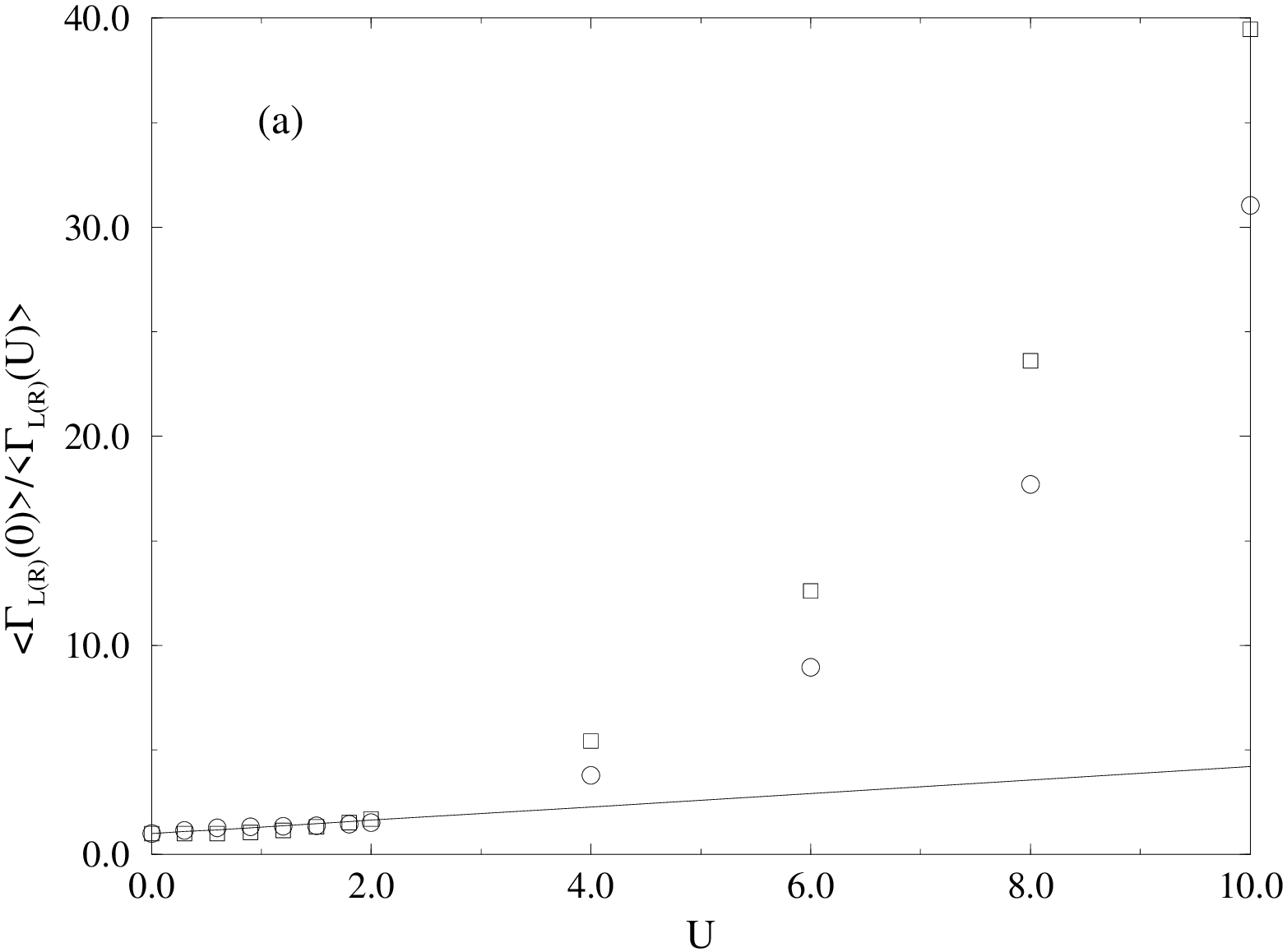}}
\vspace{-1cm}
\centerline{\epsfxsize = 3in \epsffile{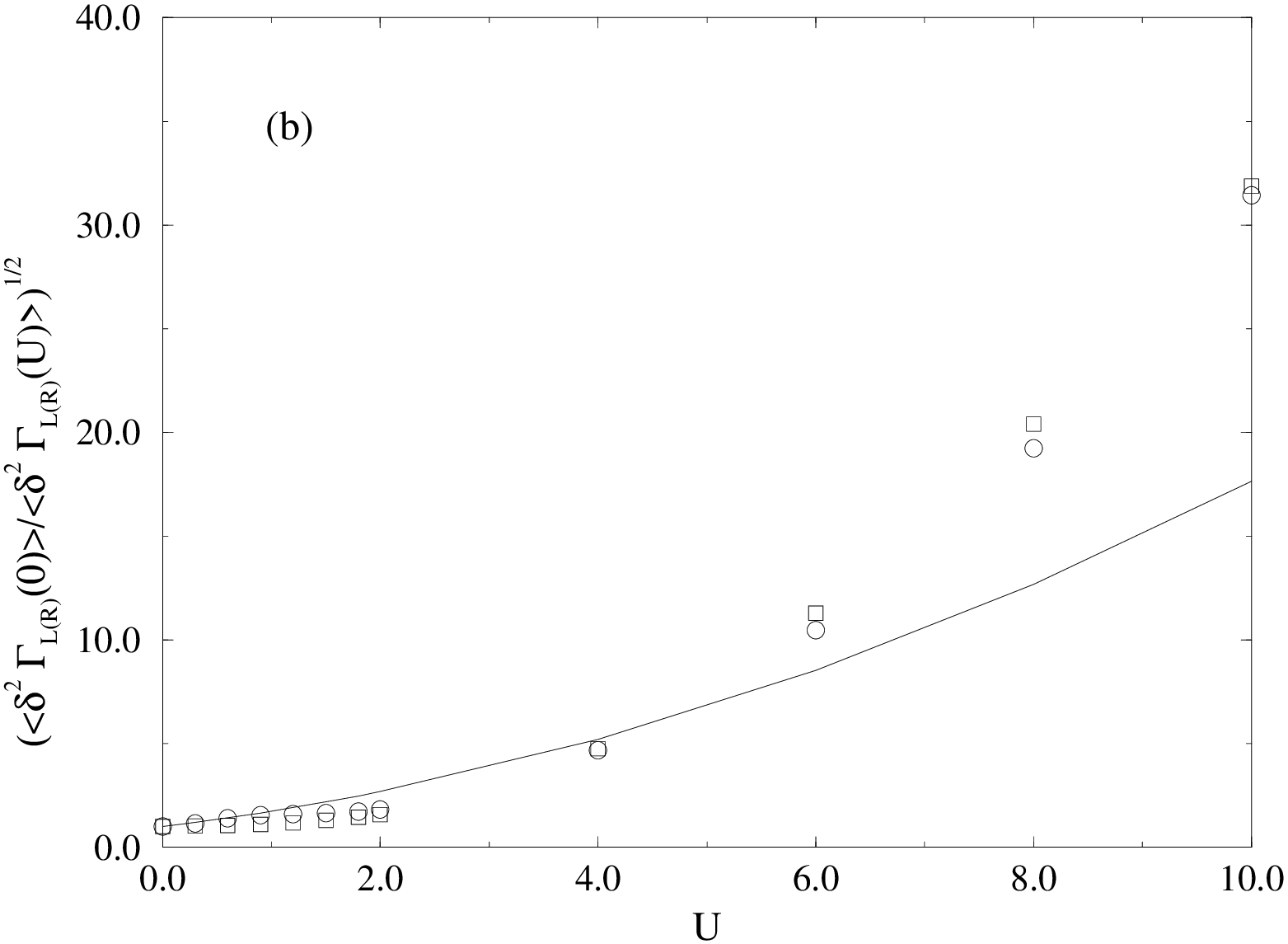}}
\caption {The influence of the e-e interaction strength on the tunneling
rate $\protect \Gamma_{L(R)}$. Circles correspond to $\protect \phi=0$, 
while squares to $\protect \phi=0.4 \phi_0$.
The lines represent the prediction of an RPA theory.
(a) The average tunneling rate, (b) the variance.
\label{fig.1}}
\end{figure}

The average tunneling rate, its root mean square value,
and the normalized fluctuations are presented in Fig \ref{fig.1}
as function of the interaction strength $U$ for $B=0$ and $B\ne0$. 
Using the expression of $\Gamma_{L (R)}$ in terms of Green functions given by
Zyuzin and Spivak \cite{zs} one can use a diagrammatic summation similar to
the one used in Fig. 1(a) of Ref. \cite{ba} for $\partial N/\partial \mu$ to
obtain the RPA predictions for $\langle\Gamma_{L(R)}(U)\rangle$.
For small
values of $U$ the RPA prediction\cite{ber}
$\langle\Gamma_{L(R)}(0)\rangle/\langle\Gamma_{L(R)}(U)\rangle
\sim 1 + (\kappa L / \pi)\sim 1 + 0.34U$ 
(where $\kappa = S_d \nu e^2$, $S_d=2 \pi$ for an infinite system and
$2.50$ for a $4\times6$ lattice\cite{ba},
$\nu=(\Delta L^2)^{-1}$ and $L^2=L_x L_y$)
is followed for both $B=0$ and $B\ne0$.
For larger values of $U$, the tunneling rate,
$\Gamma_{L(R)}(U)$ is reduced below the RPA value. It is 
expected that the short range order induced by the interactions indeed
reduce the
overlap between $a^{\dag}|\Psi_N \rangle$ and $| \Psi_{N+1} \rangle$.

The fluctuations in the tunneling rate depicted  in Fig. \ref{fig.1}b, 
also decrease
as function of the interaction strength.
In the RPA regime the fluctuations may be calculated by a diagrammatic
expansion similar to the one used in  Fig. 1(b) of Ref. \cite{ba} for
$\delta^2 \partial N/\partial \mu$. As in Refs. \cite{zs,ba} a cutoff
must be used in order to avoid divergences in the non-interacting case,
but the influence of the interaction is cutoff independent resulting in 
$(\langle\delta^2 \Gamma_{L(R)}(0)\rangle / \langle\delta^2 
\Gamma_{L(R)}(U)\rangle)^{1/2}= (1 + (\kappa L / \pi))^2$,
while for stronger interactions $\langle\delta^2 \Gamma_{L(R)}(U)\rangle$
is strongly suppressed.

To make connection to real samples we use the ratio between
the average inter-particle Coulomb interaction and the Fermi energy
$r_s=1/\sqrt{\pi n} a_B$ (where $n$ is the electronic density and
$a_B$ is the Bohr radius) corresponding to
$r_s\sim\sqrt{\pi/6}(U/2V)$ for $N=4$, $M=24$.
For all the above quantities there is a clear borderline around
$U=2-4 V$ or $r_s\sim 1$. At low $r_s$ values (high densities)
the tunneling rate agrees well with RPA calculations, while for
stronger interactions the results are qualitatively different. Identical
behavior was observed for conductance peak spacings fluctuations discussed
in Ref. \cite{sba}. In both cases the appearance of short range
correlations at $r_s \geq 1$ 
lead to a failure of RPA. It is important to bear in mind
that for real systems the density in the leads  $n = 2 - 3.5 \times 10^{11} 
\ _{\rm cm^{-2}}$, while in the dot the density is probably lower, thus
in all experimental systems 
$r_s \sim 1 - 2$. They hence correspond to
a regime where RPA no longer holds.

\begin{figure}
\vspace{-1.5cm}
\centerline{\epsfxsize = 3in \epsffile{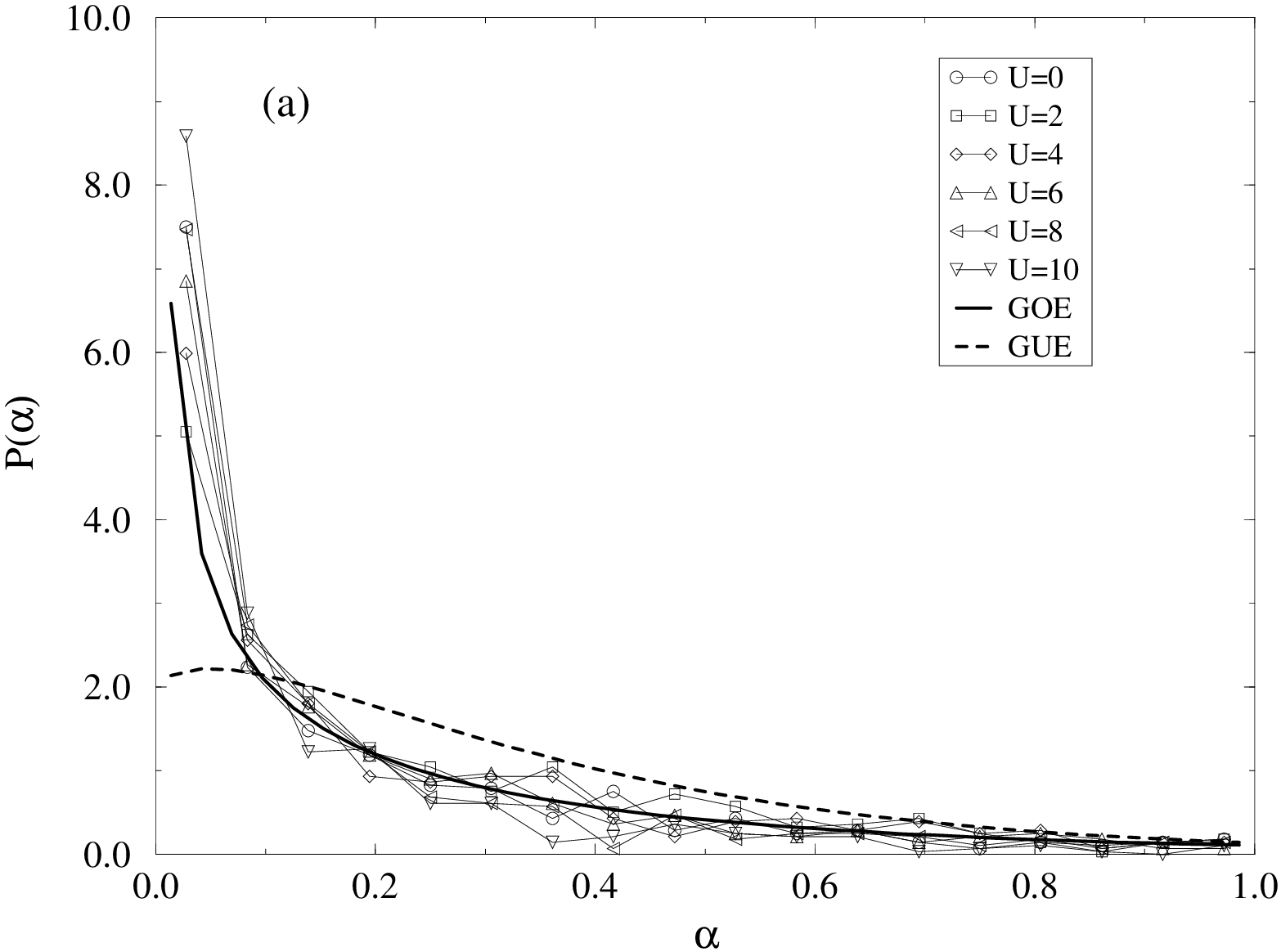}}
\vspace{-1cm}
\centerline{\epsfxsize = 3in \epsffile{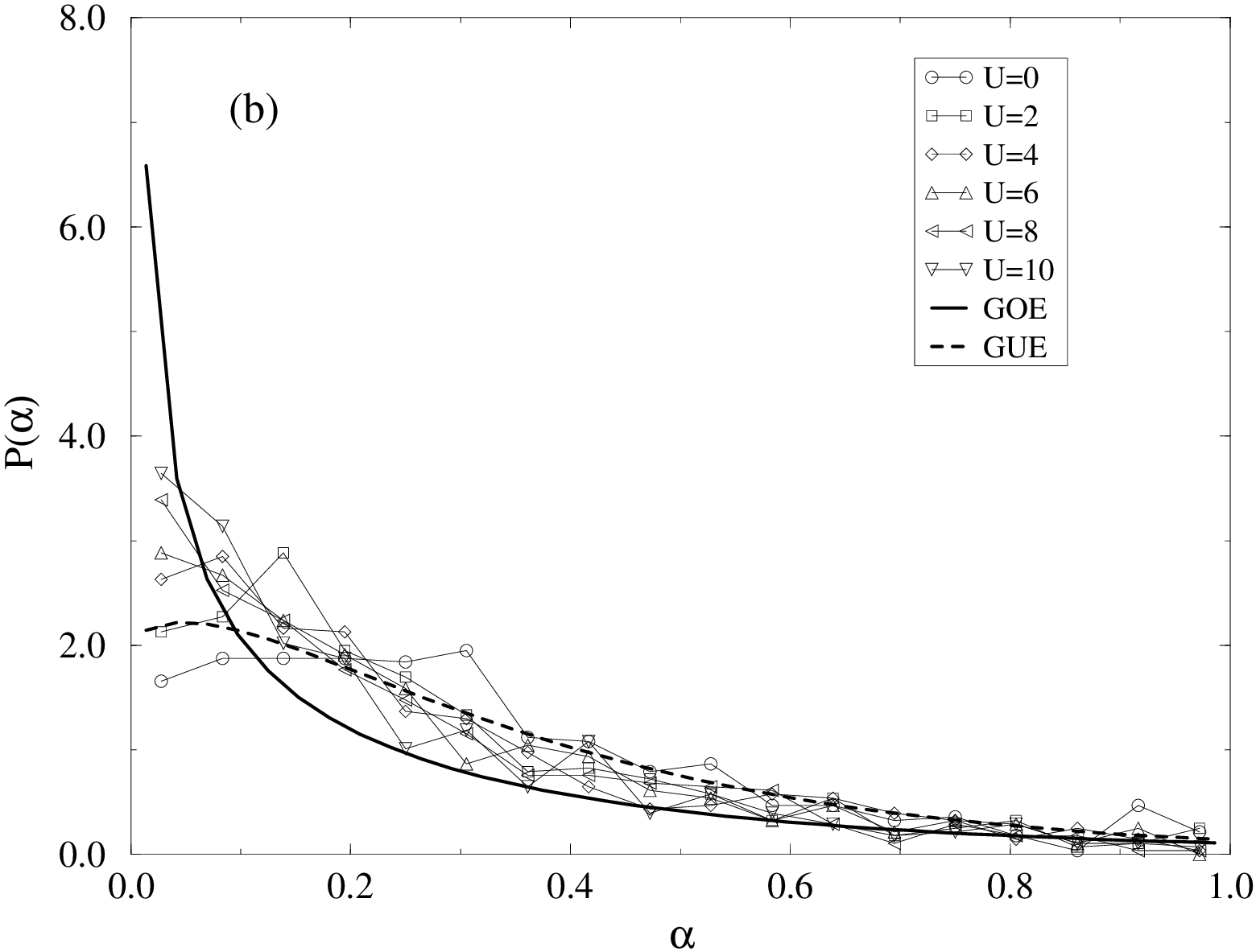}}
\caption {The probability distribution of the dimensionless 
parameter $\protect \alpha$
corresponding to the conductance peak height $\protect g_{max}$
for different values of the e-e interaction strength. (a) $\protect \phi=0$,
(b) $\protect \phi=0.4 \phi_0$. The lines correspond to the RMT predictions
for $\protect B=0$ (GOE) and $\protect B\ne0$ (GUE).
\label{fig.2}}
\end{figure}

The full probability distribution of the dimensionless parameter $\alpha$ for 
different values of the interaction is shown in Fig. \ref{fig.2}.
For $B=0$ the distributions for all values of $U$ are reasonably close
to the RMT prediction, Eq. (\ref{pa}). Thus, moderate 
changes in the second moment of $\Gamma_{L(R)}$ (Fig. \ref{fig.1}c)
hardly influence the distribution of $\alpha$.
This is in good agreement with the experimental data\cite{marc,chang}.
Also in the presence of a magnetic field, the interaction strength
has no major effect. Nevertheless, it is interesting to note 
that the dip at small values
of $\alpha$ predicted by RMT (Eq. (\ref{pa})) seen for small values
of $U$ (actually the dip is even larger than predicted which is an
artifact of the small system size) disappears for higher
values of $U$. The disappearance of the dip at higher interaction values is the
result of the reduced sensitivity of the system to the magnetic field.

The auto-correlation, Eq. (\ref{acd}), 
between the height of the i-th peak at different
values of magnetic fields is given according to RMT\cite{aa,blm} 
for the GUE ensemble by
\begin{eqnarray}
C(\phi,\Delta \phi)=
\left[1 +  \left({{\Delta \phi}\over{\phi_c}}\right)^2\right]^{-2}.
\label{ac}
\end{eqnarray}
Here $\phi=B A$ is the magnetic flux through a dot of area $A$, and
$\phi_c = \phi_0/\sqrt{g {\cal K}}$, 
where for a diffusive dot $g$ is the dimensionless
conductance $g=E_c/\Delta$ ($E_c$ is the Thouless energy)
and ${\cal K} = 1$. For a ballistic dot, 
$E_c = v_F / \sqrt{A}$, and the geometrical factor, for the case of a 
flux line threading the dot ${\cal K}$
is of order of unity\cite{blm}. This corresponds to $\phi_c \sim 0.1 \phi_0$
for the experimental setup in Ref. \cite{marc}, while the experimental
value is $\phi_c \sim \phi_0$ \cite{marc1}. Recently Alhassid \cite{al}
pointed out that for a uniform magnetic field in the dot ${\cal K}$
is much smaller corresponding to $\phi_c \sim 0.5 \phi_0$, which is still
significantly lower than the experimental value.

\begin{figure}
\centerline{\epsfxsize = 4in \epsffile{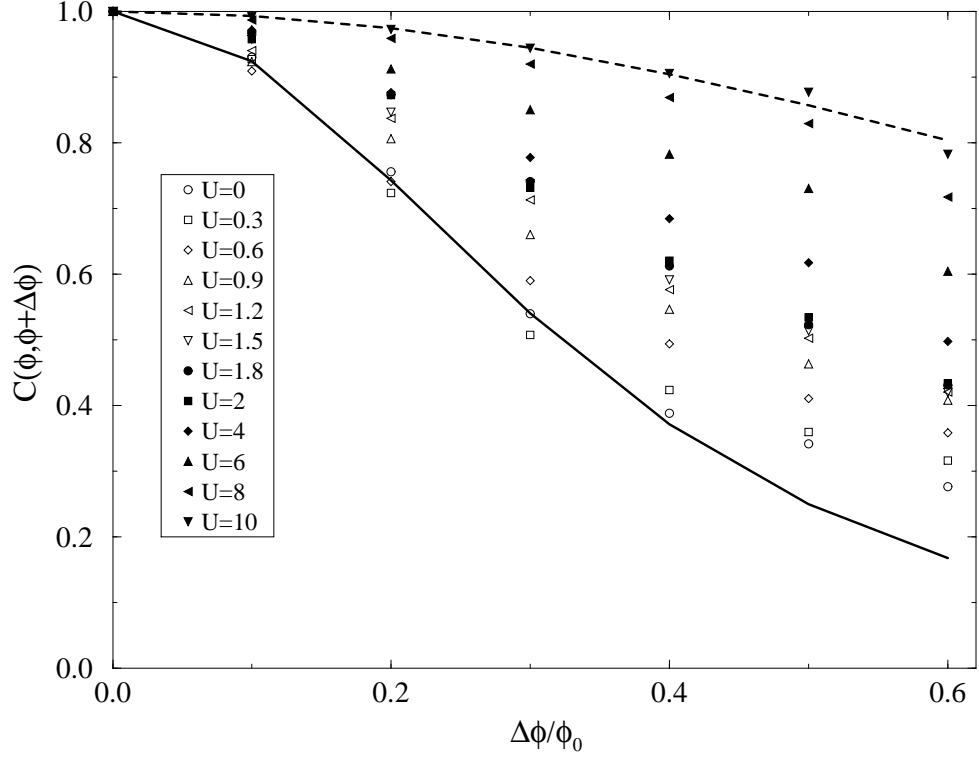}}
\caption {The auto-correlation function for different values of the e-e 
interaction strength. The full line corresponds to Eq. (\protect \ref{ac})
with $\protect \phi_c=0.5\phi_0$, while the dashed line to
Eq. (\protect \ref{ac}) with $\protect \phi_c=1.75\phi_0$.
\label{fig.3}}
\end{figure}

The numerically calculated
$C(\phi,\Delta\phi)$ for $\phi=0.2\phi_0$
for different values of $U$ is presented in Fig \ref{fig.3}.
Eq. (\ref{ac})
describes the $U=0$ behavior for small values of $\Delta\phi$ quite well.
From the value of $\phi_c=0.5\phi_0$ one obtains $g=4$ which is reasonable.

Since the  $\phi$ dependent part in the diagrammatic calculation 
(which is similar to the calculation of the fluctuations \cite{zs,ba}) does
not depend on the interaction,
one expects to first order no changes in Eq. (\ref{ac}) from RMT calculations 
\cite{ber}. Indeed, for small $U$ there is 
only a weak
influence of the interaction strength on the auto-correlation function. 
At larger values of $\Delta\phi$, $C(\phi,\Delta\phi)$
seems to be more sensitive
to $U$ and in the regime $V<U<4V$ it is hard to fit it by a
particular functional form. For $U>4V$ ($r_s\geq\sqrt2$)
the auto-correlation function can be fitted 
again to the functional form of Eq. (\ref{ac}) but the
correlation flux is enhanced. For example, for $U=10V$ $\phi_c=1.75\phi_0$.
Thus, although strong interactions reduce the average conductance
peak height, they hardly influence the conductance distribution,
again in agreement with both experiments \cite{marc,chang}.
The main effect is a reduced sensitivity to flux which is also manifested in
the weak dependence of the conductance peak spacing
fluctuations\cite{shw} on magnetic field. It is
worth mentioning that tunneling experiments through excited states of
heavily doped
GaAs dots \cite{Sivan94}, where $r_s<1$, find the RMT correlation flux.
The numerical calculation gives the two particle correlation function as 
well\cite{ba}. It turns out that the point where the conductance starts
to deviate from RPA ($r_s \sim 1$) is accompanied by the appearance of
short range correlations. the same correlations that led to the large 
conductance peak spacing 
fluctuations are here responsible for the transmission 
statistics. We emphasize that Wigner crystallization occurs at much
stronger interactions.

In conclusion, the appearance of short range electron correlations
which are the result of
the low electronic densities in the measured quantum dots explains
not only the large conductance peak spacing fluctuations, but also other 
features pertaining to their transport properties. 
First, it has been numerically
shown that although the conductance peak hight distribution
is well described by the 
single electron RMT,
this distribution is valid also at the low density (strong interaction) 
regime. The strong interaction model also agrees with the results
of the experiment in the presence of a magnetic field, which stands
at odds with the predictions of the single electron RMT. Most importantly, 
the auto-correlation between
peak heights exhibits an enhanced characteristic 
magnetic field for $r_s>1$ which is
in good agreement with the experimental observations\cite{marc,chang,marc1}.
Thus, the fact that the experiments are performed at low densities ($r_s>1$)
for which the RPA theory no longer holds explains
many of the puzzling features exhibited by these systems. 
A unified explanation for the large conductance peak spacing fluctuations
and the peak height characteristics is hence provided. A better
understanding of temperature effects, and a full analytic treatment of
the short range correlations is still needed.

Useful discussions with A. Chang and C. Marcus are gratefully acknowledged.
R.B. would like to thank the Minerva Center for the Physics of Mesoscopics, 
Fractals and Neural Networks 
and  the US-Israel Binational Science Foundation 
for financial support.
U.S. would like to thank the US-Israel Binational Science Foundation 
and the Technion grant for the promotion of research for financial support.

\end{document}